\documentclass[12pt,english]{article}
\usepackage[T1]{fontenc}
\usepackage[latin2,latin1]{inputenc}
\usepackage[colorlinks,ps2pdf]{hyperref}

\makeatletter

\providecommand{\tabularnewline}{\\}

\usepackage{graphicx}

\usepackage{babel}
\makeatother
\begin{document}

\title{\textcolor{blue}{\bf The compact star in the SU(3) sigma model}}

\author{\textcolor{red}{\bf Ryszard Manka, Ilona Bednarek}}

\date{University of Silesia, Katowice, Poland.}

\maketitle
\begin{abstract}
\textcolor{red}{The linear chiral $SU(3)_{L}\times SU(3)_{R}$ model is applied to
describe properties of the compact star matter inside the quark, protoneutron
and neutron star.} 
\end{abstract}

\section{\textcolor{blue}{Introduction}}

In this presentation we shall investigate the astrophysical properties
of the compact stars (protoneutron, neutron and quark stars \cite{blen})
within the Relativistic Mean Field (RMF) \cite{walecka} theory originated
from the linear SU(3) sigma model \cite{pap1,pap2,torn}.

The main aim of this work is to show how an effective mean field approximation
(RMF) emerges from the linear SU(3) chiral model and to show comparison
to the current RMF approach (Furnstahl - Serot - Tang (FST) model
\cite{serot}) as well as to present astrophysical applications in
the form of the neutron star model. The effective model which includes
scalar, vector and scalar-vector interaction terms is applied to describe
properties of the quark, protoneutron and neutron star matter.

\section{\textcolor{blue}{The SU(3) sigma model}}

\begin{figure}
  \begin{center}
  \includegraphics[width=8cm]{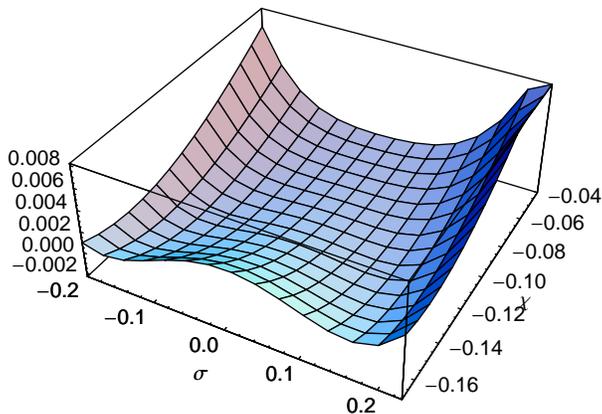}
  \end{center}
\caption{The effective potential $U_{eff}(\sigma,\chi)$ 
in the SU(3) sigma model.}
\label{cap:1a}
\end{figure}

The chiral SU(3) model was proposed by Papazoglou \emph{et al.} \cite{pap1,pap2}.
In the original form it describes interaction of the baryons and mesons
SU(3) multiplets. Recently, a chiral SU(3) quark model has been proposed
by Wang \emph{et al} \cite{wang}. The basic fields that compose the
theory represent the realization of the group $SU(3)_{L}\times SU(3)_{R}$.
The meson content of the model is scalar, pseudo-scalar and vector.
Naive quark models interpret them as exited $\overline{q}q$ states.
Scalar and pseudo-scalar mesons can be grouped into 
\begin{equation}
\Phi=\Sigma+i\Pi=\frac{1}{\sqrt{2}}T_{a}\phi_{a}=\frac{1}{\sqrt{2}}T_{a}(\sigma_{a}+i\pi_{a})
\end{equation}
 where $\sigma_{a}$ and $\pi_{a}$ are members of the scalar and
pseudo-scalar octet respectively: 
\begin{equation}
\Sigma=\frac{1}{\sqrt{2}}\sigma^{a}\lambda^{a}=\left\{ \begin{array}{ccc}
\frac{1}{\sqrt{2}}(f_{0}+a_{0}^{0}), & a_{0}^{+}, & K^{+}\\
a_{0}^{-}, & \frac{1}{\sqrt{2}}(f_{0}-a_{0}^{0}), & K^{0}\\
K^{-}, & \bar{K}^{0}, & f'_{0}\end{array}\right\} 
\end{equation}
\begin{equation}
\Pi=\frac{1}{\sqrt{2}}\pi^{a}\lambda^{a}=\left(\begin{array}{ccc}
\frac{1}{\sqrt{2}}(\pi^{0}+\eta^{8}) & \pi^{+} & \kappa^{+}\\
\pi^{-} & \frac{1}{\sqrt{2}}(-\pi^{0}+\eta^{8}) & \kappa^{0}\\
\kappa^{-} & \kappa^{0} & \zeta\\
\end{array}\right).
\end{equation}
 The vector meson octet is given by 
 \begin{equation}
V=\frac{1}{\sqrt{2}}v^{a}\lambda^{a}=\left(\begin{array}{ccc}
\frac{1}{\sqrt{2}}(\omega+\rho_{0}^{0}), & \rho_{0}^{+}, & K^{\ast+}\\
\rho_{0}^{-}, & \frac{1}{\sqrt{2}}(\omega-\rho_{0}^{0}), & K^{\ast0}\\
K^{\ast-}, & \overline{K^{\ast0}}, & \phi\\
\end{array}\right).
\end{equation}
The most general form of the Lagrangian function can be written as
a sum of the following parts 
\begin{equation}
\mathcal{L}=\mathcal{L}_{M}+\mathcal{L}_{F}+\mathcal{L}_{sb}
\end{equation}
where 
 \begin{equation}
\mathcal{L}_{M}=\frac{1}{2}Tr(\partial_{\mu}\Phi\partial^{\mu}\Phi)-\frac{1}{2}\mu^{2}Tr(\Phi^{2})-\frac{\lambda}{4}Tr((\Phi^{+}\Phi)^{2})-\frac{\kappa}{4}Tr(\Phi^{+}\Phi)^{2}\end{equation}
is the Lagrangian function which describes scalar and pseudo-scalar
mesons $\Phi$ $(\Phi=\Sigma+i\Pi)$. The symmetry breaking term $\mathcal{L}_{sb}$
has the form of: 
\begin{equation}
\mathcal{L}_{sb}=\frac{1}{2}c(Det(\Phi)+Det(\Phi)^{\ast})+Tr(H^{+}\Phi+\Phi^{+}H).
\end{equation}
In the mean field approximation the chiral symmetry is broken and
the meson fields gain non-vanishing vacuum expectation values ($\sigma,\chi$)
\[
<\Phi>=<\Sigma>=\left \{ 
\begin{array}{ccc}
\frac{1}{\sqrt{2}}\sigma, & 0, & 0\\
a_{0}^{-}, & \frac{1}{\sqrt{2}}\sigma, & 0\\
0, & 0, & \chi
\end{array}\right\} .
\]
The effective potential $U_{eff}(\sigma,\chi)=-<L>$ (Fig. \ref{cap:1a})
determines the scale of the chiral symmetry breaking. %
Shifting the meson field $\Phi=\bar{\Phi}+<\Phi$> ($f_{0}=\bar{f}_{0}+\sigma$,
$f'_{0}=\bar{f}'_{0}+\chi$) and diagonalizing the square mass matrics
$m_{a,b}^{2}=\frac{\partial^{2}U_{eff}}{\partial\bar{\sigma}_{a}\partial\bar{\sigma}_{b}}$
produce the physical meson fields 
\begin{equation}
\left\{ \begin{array}{c}
\varphi\\
\varphi_{*}\end{array}\right\} =\left(\begin{array}{cc}
cos\vartheta & sin\vartheta\\
-sin\vartheta & cos\vartheta\end{array}\right)\left\{ \begin{array}{c}
f_{0}\\
f'_{0}\end{array}\right\} .\label{eq:fx}
\end{equation}
 Fitting the model to the observed masses of mesons allows to determine
its parameters (similar to the case \cite{rischke} of the explicit
chiral symmetry breaking with $U(1)_{A}$ anomaly). It gives $\mu=552.274\, MeV$,
$\lambda=45.11$, $\kappa=-8.917$ and $c=3412.25$. This fitting
gives meson masses (e.g. $\pi$, $K$, $\varphi,$ $\varphi_{*}$,
$\delta=a_{0}$, etc.) including sigma meson mass $m_{\sigma}=502.27\, MeV$.
The only unclear thing is the sigma meson mass. The light scalar meson
$\sigma$ (denoted here as $\varphi$) is an elusive subject of classification.

The fermion contents of the model consists of either quarks or baryons,
respectively (QMF or RMF model).

The chiral SU(3) quark mean field model has been applied to describe
the quark matter or nucleon matter. In the chiral limit, the quark
field $q=\{ u,d,s\}$ with three flavors, can be decomposed into left
and right-handed parts $q=q_{L}+q_{R}$. The quarks are described
by the Lagrange function
\[
\mathcal{L}_{F}=i\bar{q}\gamma^{\mu}D_{\mu}q-\bar{q}m_{0}q+g_{s}\bar{q}\Phi q-\chi_{c}(r)\bar{q}q
\]
 where $\chi_{c}(r)$ is quarks confining potential. Solving the Dirac
equation for quark in confining potential the baryon masses can be
calculated 
\[
M_{eff,N}(\varphi,\varphi_{*})=M_{N}-g_{\sigma}(\varphi)\,\varphi-g_{\sigma*}\varphi_{*}=M_{N}-g_{\sigma}\,\varphi+\frac{1}{2}g_{\sigma}C^{\prime}(0)\,\varphi^{2}+...
\]
with $g_{\sigma}(\varphi)=g_{\sigma}-\frac{1}{2}g_{\sigma}C^{\prime}(0)\,\varphi=g_{\sigma}-\frac{1}{2}a\,\varphi$.
The last nonlinear term indicates the inner nucleon structure.

\section{\textcolor{blue}{The effective RMF approach}}

\begin{figure}
\includegraphics[%
  width=8cm]{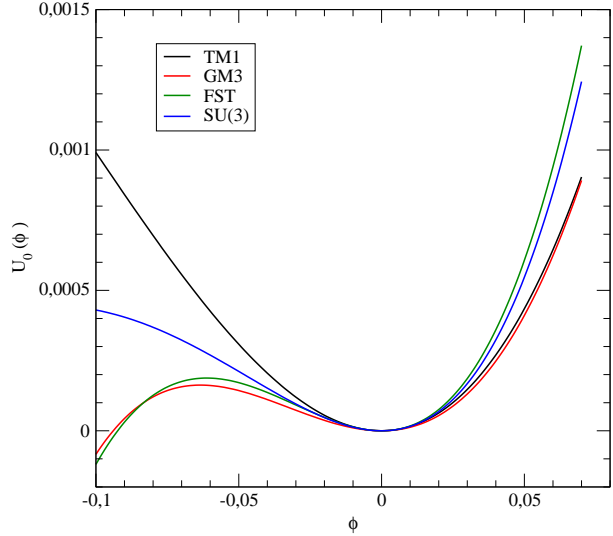}
  \centering

\caption{The potential $U_{0}$ for the scalar meson $\varphi$ of the effective
RMF theory.\label{cap:2b}}
\end{figure}

The nuclear relativistic mean field approach describes the nuclear
interactions due to the mesons exchange between baryons ($p,$~$n$,~$\Lambda,\,\Sigma,\,\Xi$).
Baryons are grouped into the isospin and hipercharge representations
$(\frac{1}{2},1)$, $(\frac{1}{2},-1)$, $(1,0)$
\begin{equation}
\Lambda,\,\, N=\left(\begin{array}{c}
p\\
n\end{array}\right),\,\,\Xi=\left(\begin{array}{c}
\Xi^{0}\\
\Xi^{-}\end{array}\right),\,\,\Sigma=\left(\begin{array}{c}
\Sigma^{+}\\
\Sigma^{0}\\
\Sigma^{-}\end{array}\right).\end{equation}
 The RMF Lagrange functions \begin{equation}
\mathcal{L}_{RMF}=\mathcal{L}_{B}+\mathcal{L}_{M}\end{equation}
 describes baryons ($B=\{ b,n,\Lambda,\Sigma,\Xi\}$) \begin{equation}
\mathcal{L}_{B}=\sum_{B}i\overline{\psi}_{B}\gamma^{\mu}D_{\mu}\psi_{B}-\sum_{B}M_{B}(\varphi,\varphi_{*})\overline{\psi}_{B}\psi_{B}\end{equation}
 and mesons \begin{equation}
\mathcal{L}_{M}=\mathcal{L}_{Ms}+\mathcal{L}_{Mv}.\end{equation}
 Mesons can be divided into scalar mesons ($\varphi,$ $\varphi_{*},$
$\delta$) described by $\mathcal{L}_{Ms}$ and vector mesons ($\omega$,
$\rho$, $\phi$) described by $\mathcal{L}_{Mv}$ . \begin{eqnarray}
 & \mathcal{L}_{Ms}=\frac{1}{2}\partial_{\mu}\varphi\partial^{\mu}\varphi+\frac{1}{2}\partial_{\mu}\delta^{a}\partial^{\mu}\delta^{a}++\nonumber \\
 & \frac{1}{2}\partial_{\mu}\varphi_{*}\partial^{\mu}\varphi_{*}-U_{S,eff}(\varphi,\varphi_{*},\delta),\end{eqnarray}
\begin{eqnarray}
 & \mathcal{L}_{Mw}=-\frac{1}{4}\Omega_{\mu\nu}\Omega^{\mu\nu}+\frac{1}{2}m_{\omega}^{2}\omega_{\mu}\omega^{\mu}-\frac{1}{4}R_{\mu\nu}^{a}R^{a\mu\nu}+\frac{1}{2}m_{\rho}^{2}\rho_{\mu}^{a}\rho^{a\mu}\nonumber \\
 & -\frac{1}{4}\Phi_{\mu\nu}\Phi^{\mu\nu}+\frac{1}{2}m_{\phi}^{2}\phi_{\mu}\phi^{\mu}+U_{V,eff}(\omega,\rho)\end{eqnarray}
 where \begin{eqnarray}
 & \Omega_{\mu\nu}=\partial_{\mu}\omega_{\nu}-\partial_{\nu}\omega_{\mu}\,\,\,\,\,,\,\,\,\,\,\,\Phi_{\mu\nu}=\partial_{\mu}\phi_{\nu}-\partial_{\nu}\phi_{\mu}\\
 & R_{\mu\nu}=R_{\mu\nu}^{a}T^{a}=\partial_{\mu}R_{\nu}-\partial_{\nu}R_{\mu}-ig_{\rho}[R_{\mu},R_{\nu}].\end{eqnarray}
 The scalar meson interaction is enormously nonlinear. It comes from
the shifting in the potential $U_{eff}(\sigma,\chi)$ according to
the prescription (eq. \ref{eq:fx}). In the simplest approximation
this procedure generates the polynomial scalar interaction (Fig. \ref{cap:2b})
of the RMF approach 
\begin{equation}
U_{0}(\varphi)=U_{eff}(\sigma_{0}+\varphi,\chi_{0})=U_{eff}(\varphi,0)=\frac{1}{2}m_{\sigma}^{2}\varphi^{2}+\frac{1}{3}g_{2}\varphi^{3}+\frac{1}{4}g_{3}\varphi^{4}\end{equation}

\begin{figure}
\begin{center}\includegraphics[%
  width=8cm]{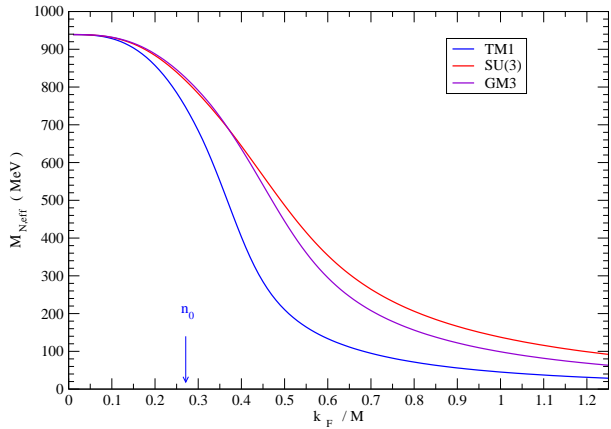}\end{center}

\caption{The effective nucleon masses for different parameter sets as a function
of the nucleon Fermi momentum $k_{F}$.}

\label{meff}
\end{figure}

The form of the potential function was first introduced by Boguta
and Bodmer \cite{bodner} in order to get the correct value of the
compressibility $K$ of nuclear matter at saturation density (see
Table 1). The simplest Walecka model (L2) (\emph{linear} Walecka model
$g_{2}=g_{3}=0$) brings a very large, unrealistic value of the parameter
$K$ \cite{walecka}. Fig. \ref{meff} depicts the effective nucleon
masses obtained for different parameter sets functions of baryon number
density $n_{B}$. The parameters describing the nucleon-nucleon interactions
in the RMF approach are chosen in order to reproduce the properties
of the symmetric nuclear matter at saturation such as the binding
energy, symmetry energy and incompressibility. In the chiral $SU(3)$
model they are generally calculable from the starting ones ($\mu^{2}$,
$\lambda$, $\kappa$) which are fitted to the mesons spectroscopy.
 The appropriate parameter set is constrained not only by the value
of physical scalar meson masses but also by the properties of nuclear
matter at saturation. For symmetric nuclear matter the nucleon density
equals ~$n_{0}=2.5\,10^{14}\,$ $g\, cm^{-3}$ $=0.15\, nucleons/fm^{3}$
$=140$~ MeV$\, fm^{-3}$. The obtained results are collected in
Table 1.%
\begin{figure}
\begin{center}\includegraphics[%
width=8cm]{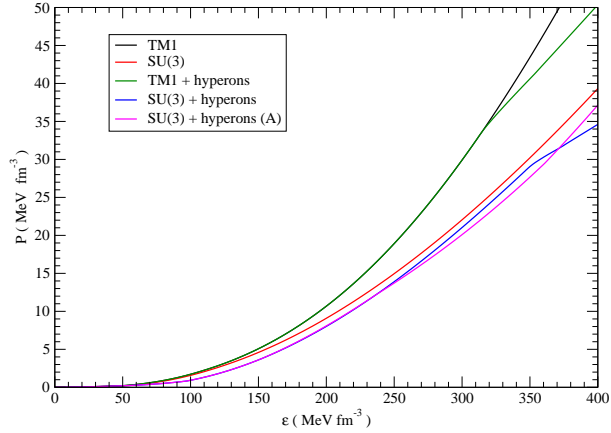}\end{center}

\caption{The nuclear matter equation of state for different parameter sets.}

\label{eos}
\end{figure}
\begin{table}
\begin{tabular}{|c|c|c|c|c|}
\hline 
Parameter&
 GM3\cite{glen}&
 TM1\cite{tm1}&
 FST\cite{serot}&
 SU(3) \tabularnewline
\hline
$E_{0}$~(MeV)&
 -16.35&
 -16.26&
 -16.38 &
 -16.31 \tabularnewline
\hline
$\delta_{0}$&
 0.793 &
 0.659 &
 0.661&
 0.761\tabularnewline
\hline
$n_{0}$~$(fm^{-3})$&
 0.153 &
 0.145&
 0.155&
 0.145\tabularnewline
\hline
$K$~(MeV)&
 241.12 &
 281.53&
 219.5&
 194.9\tabularnewline
\hline
$J$~(MeV) &
 32.44 &
 36.82&
 38.17&
 34.70  \tabularnewline
\hline
\end{tabular}\centering

\caption{Properties of the nuclear matter at saturation for the symmetric
nuclear matter.}
\end{table}

\section{\textcolor{blue}{The compact star in the SU(3) sigma model}}

The compact star is a result of the equilibrium between gravitational
collapse and the pressure generated by the nuclear or quark matter
and leptons \cite{pras, kos}. With respect to the electroweak interaction the matter
is in $\beta$ equilibrium. In the neutron star weak interactions
are responsible for $\beta$ decay 
\begin{equation}
n+\nu_{e}\,\leftrightarrow\, p+e,
\end{equation}
\begin{equation}
\mu+\nu_{e}\,\leftrightarrow\, e+\nu_{\mu}.
\end{equation}
These reactions produce appropriate relation among the chemical potentials
of neutrinos in a protoneutron star where they are trapped 
\begin{eqnarray}
 & \mu_{p}=\mu_{n}+\mu_{\nu_{e}}-\mu_{e}\\
 & \mu_{\nu_{e}}=\mu_{e}+\mu_{p}-\mu_{n}\\
 & \mu_{\nu_{\mu}}=\mu_{\nu_{e}}-\mu_{e}+\mu_{\mu} & .
\end{eqnarray}

\subsection{\textcolor{green}{The protoneutron star.} }

Protoneutron stars  are hot and lepton rich objects formed as a
result of type II supernovae explosion.The collapse of an iron core of a massive star leads to the
formation of a core residue which is considered as an intermediate
stage before the formation of a cold, compact neutron star. This
intermediate stage which is called a protoneutron star can be
described \cite{pra1,pra2,pra3,bur} as
 a hot, neutrino opaque core, surrounded by a colder neutrino
transparent outer envelope. The evolution of a nascent neutron
star  can be described by a series of separate phases starting
from the moment when the star becomes gravitationally decoupled
from the expanding ejects. In this paper two evolutionary phases
which can be characterized by the following assumptions:
\begin{itemize}
    \item the low entropy core  $s/n_B=1-2$ (in units of the Boltzmann's constant) with trapped
    neutrinos $Y_L=0.4$
    \item the cold, deleptonized core ($Y_L=0, s/n_B=0$).
\end{itemize}
 have been considered.
These two distinct stages are separated by the period of
deleptonization. During this epoch the neutrino fraction decreases
from the nonzero initial value ($Y_{\nu}\neq 0$) which is
established by the requirement of the fixed  total lepton number
at $Y_l=0.4$,  to the final one characterized by $Y_{\nu}=0$.
Evolution of a protoneutron star which proceeds by neutrino
emission causes that the star changes itself from a hot, bloated
object to a cold, compact neutron star.
\newline
The interior of this  very early stage of a protoneutron star is
an environment in which  matter with the value of entropy of the
order of 2 with trapped neutrinos produces a pressure to oppose
gravitational collapse. The lepton composition of  matter is
specified by the fixed lepton number $Y_L=0.4$. Conditions that
are indispensable for the unique determination of the equilibrium
composition of a protoneutron star matter  arise from the
requirement of $\beta$ equilibrium, charge neutrality and baryon
and lepton number conservation. The later one is strictly
connected with the assumption that the net neutrino fraction
$Y_{\nu}\neq 0$ and therefore the neutrino chemical potential
$\mu_{\nu} \neq 0$. When the electron chemical potential $\mu_e$
reaches the value equal to the muon mass, muons start to appear.

\begin{figure}
  \begin{center}\includegraphics[%
    width=8cm]{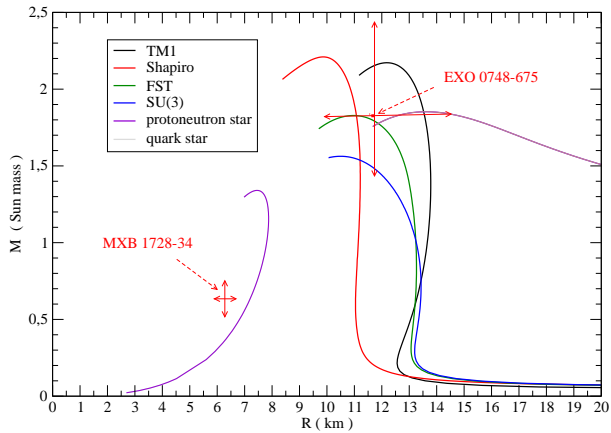}
  \end{center}

\caption{The mass radius relation for various quark and neutron stars.}

\label{rm}
\end{figure}

In a protoneutron star, when the neutrino is trapped, its chemical
potential strongly depends on nuclear asymmetry 
\begin{equation}
\mu_{\nu_{e}}=\mu_{e}-\mu_{A},
\end{equation}
\begin{equation}
\mu_{A}=\mu_{n}-\mu_{p}=\varepsilon_{n}-\varepsilon_{p}-g_{\rho}r_{0}
\end{equation}
\begin{equation}
\epsilon_{p,n}=\sqrt{k_{F,B}^{2}+M_{B,eff}^{2}}|_{B=n,p}.
\end{equation}
where $r_{0}=<\rho_{0}^{3}>$ is the expected value for the $\rho$
meson in medium. If $\nu_{\mu}$ are able to escape from the protoneutron
star and $\mu_{e}$ not, the muon chemical potential \[
\mu_{\mu}=\mu_{A}\]
depends on the nuclear asymmetry. In conclusion, the protoneutron
star with the biggest neutrino number is in the nuclear symmetric
phase, when $\mu_{A}=0$, so $\mu_{\nu_{e}}=\mu_{e}$ and muons are
absent $\mu_{\mu}=0$. The neutrinos as fermions increase the star
pressure and make the star with a large radius ($\sim20-50\, km$
, Fig. \ref{rm}). It is interesting that the outer layers when density
drops below $\sim10^{14}\, g/cm^{3}$ and nuclear interaction vanishes
the weak interactions with $\beta$ decay still takes place as long
as neutrinos are trapped. The outer layers of the protoneutron star
has mainly the electroweak nature. The star core is similar to the
symmetric hot nuclear matter. The entropy per baryon $s/n_{B}=1$
gives temperature ($\sim\,\,40-80\,\, MeV$) inside the star.

\subsection{\textcolor{green}{The neutron star.}}

When neutrinos escape at last ($\mu_{\nu_{e}}=0$), the new equilibrium
have to take place at least in the neutron star case. 
Neutrinos are now neglected here since they leak out from the neutron star, whose
energy diminishes at the same time. The pressure
decrease reduces the star radius till $\sim10-14\, km$ and increases
the star density to $\sim10^{14}-10^{15}\, g/cm^{3}$, the nuclear
interactions became more crucial. The equilibrium conditions with
respect to the $\beta$ decay between baryonic (including hyperons)
and leptonic species lead to the following relations among their chemical
potentials and constrain the species fraction in the star interior
\begin{eqnarray}
\mu_{p}=\mu_{\Sigma^{+}}=\mu_{n}-\mu_{e}\hspace{10mm}\mu_{\Lambda}=\mu_{\Sigma^{0}}=\mu_{\Xi^{0}}=\mu_{n}\\
\mu_{\Sigma^{-}}=\mu_{\Xi^{-}}=\mu_{n}+\mu_{e}\hspace{10mm}\mu_{\mu}=\mu_{e}
\end{eqnarray}
The higher the density the greater number of hyperon species are
expected to appear. They can be formed both in leptonic and baryonic
processes. In the latter one the strong interaction process such as
\begin{equation}
n+n\rightarrow n+\Lambda
\end{equation}
proceeds. There are other relevant strong reactions that establish
the hadron population in neutron star matter e.g.: 
\begin{equation}
\Lambda+n\rightarrow\Sigma^{-}+p\hspace{5mm}\Lambda+\Lambda\rightarrow\Xi^{-}+p
\end{equation}
\begin{figure}
\begin{center}
 \includegraphics[width=8cm]{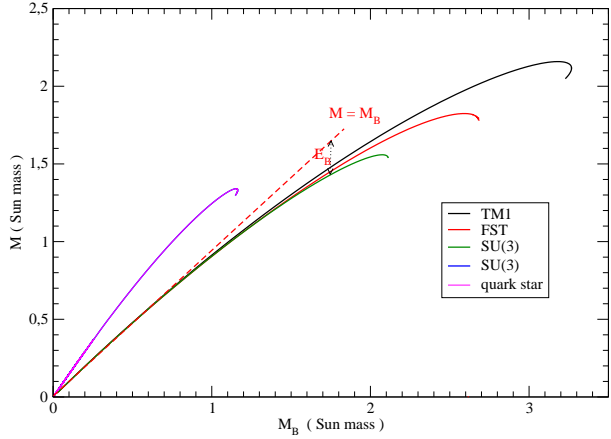}
\end{center}

\caption{The gravitational binding energy for quark and neutron stars.}

\label{grav}
\end{figure}
The final result is the equation of the state (Fig. \ref{eos}).
All these, lead to the neutron star model with the value of maximum
mass close to $1.5\,\, M_{\odot}$ with the reduced value of proton
fraction and very compact hyperon core. The FST equqtion of state gives more 
massive neutron star $\sim 1.7 \  - \  1.8 \ M_{\odot} $.

\subsection{\textcolor{green}{The quark star.}}

Quark strange stars are astrophysical compact objects which are entirely
made of deconfined \textit{u,d,s} quark matter (\textit{strange matter})
staying in  \( \beta  \) - equilibrium. The possible existence of
strange stars is a direct consequence of the conjecture
that strange matter may be the absolute ground state of strongly interacting
matter. 

Some years ago Guichon proposed an interesting model concerning  the change
of the nucleon properties in nuclear matter (quark-meson coupling
model (QMC)) \cite{aust}. The model construction mimics the relativistic
mean field theory, where the scalar \( \sigma  \) and the vector
meson \( \omega  \) fields couple not with nucleons but directly
with quarks. The quark mass has to change from its bare current mass
due to the coupling to the \textbf{\( \sigma  \)} meson. More recently,
Shen and Toki \cite{toki1} have proposed a new version of the QMC
model, where the interaction takes place between constituent quarks
and mesons. They refer the model as the quark mean field model (QMF).
In this work we shall also investigate the quark matter within the
QMF theory motivated by parameters coming from the SU(3) chiral model
The quark phase consists the consistuent quarks with three flavors
$u$, $d$ and $s$. Quarks as effectively free quasiparticles in vacuum with non-vanishing
bag 'constant'. Quarks and electrons are in $\beta$-equilibrium
which can be described as a relation among their chemical potentials
\[
\mu_{d}=\mu_{u}+\mu_{e}=\mu_{s}
\]
where $\mu_{u}$, $\mu_{d}$, $\mu_{s}$ and $\mu_{e}$ stand for
quarks and electron chemical potentials respectively. If the electron
Fermi energy is high enough (greater then the muon mass) in the neutron
star matter muons start to appear as a result of the following reaction
\begin{eqnarray*}
d\rightarrow u+e^{-}+\overline{\nu}_{e}\\
s\rightarrow u+\mu^{-}+\overline{\nu}_{\mu}\end{eqnarray*}
The neutron chemical potential is 
\[
\mu_{n}\equiv\mu_{u}+2\mu_{d}.
\]
In a pure quark state the star should to be charge neutral. This
gives us an additional constraint on the chemical potentials 
\begin{equation}
\frac{2}{3}n_{u}-\frac{1}{3}n_{d}-\frac{1}{3}n_{s}-n_{e}=0.
\end{equation}
where $n_{f}$ ($f\in u,d,s$), $n_{e}$ the particle densities of
the quarks and the electrons, respectively. The EOS can now be parameterized
by only one chemical potential, say $\mu_{u}$. 
Nowadays the strange quark star is the subject of considerable interest \cite{njp,bub,weber}.

\subsection{\textcolor{green}{The Tolman - Oppenheimer - Volkoff equations.}}

The spherically symmetric static star in general gravity is solution
of the Einstein equations 
\[
R_{\mu\nu}-\frac{1}{2}g_{\mu\nu}R=\kappa T_{\mu\nu}
\]
for the metric\[
g_{\mu\nu}=\left(\begin{array}{cccc}
-e^{\nu(r)} & 0 & 0 & 0\\
0 & e^{\lambda(r)} & 0 & 0\\
0 & 0 & r^{2} & 0\\
0 & 0 & 0 & r^{2}sin^{2}\theta\end{array}\right)\]
where $\kappa=\frac{8\pi G}{c^{4}}$ and 
\[
T_{\mu\nu}=(P+\epsilon)u_{\mu}u_{\nu}-Pg_{\mu\nu}=
\left(
   \begin{array}{cccc}
   \epsilon=c^{2}\rho & 0 & 0 & 0\\
0 & P & 0 & 0\\
0 & 0 & P & 0\\
0 & 0 & 0 & P
   \end{array}
\right)
\]
is the energy-momentum tensor describing an equation of state of the
matter. One of the Einstein equations gives
\[
e^{-\lambda}=1-\frac{2Gm(r)}{r}
\]
where $m(r)$ is the mass accumulated inside a sphere of radius, so
\begin{equation}
\frac{dm(r)}{dr}=4\pi r^{2}\rho.
\end{equation}
The continuity equation
\[
T_{\,\,\,;\nu}^{\mu\nu}=0\Rightarrow\frac{d\nu(r)}{dr}=-\frac{2}{P(r)+c^{2}\rho(r)}\frac{dP(r)}{dr}.
\]
The second Einstein equation gives the Tolman -Oppenheimer-Volkoff
equation
\begin{equation}
\frac{dP(r)}{dr}=-\frac{Gm(r)\rho(r)}{r^{2}}\frac{(1+\frac{P(r)}{\rho(r)})(1+\frac{4\pi r^{3}P(r)}{m(r)})}{1-\frac{2Gm(r)}{r}}.
\end{equation}
The obtained form of the equation of state serves as an input to
the Tolman - Oppenheimer - Volkoff equations and determines the structure of
spherically symmetric stars. The gravitational binding energy of a
relativistic star is defined as a difference between its gravitational
and baryon masses 
\begin{equation}
E_{b,g}=(M_{p}-m(R))c^{2}
\end{equation}
 where 
\begin{equation}
M_{p}=4\pi\int_{0}^{R}drr^{2}(1-\frac{2Gm(r)}{c^{2}r})^{-\frac{1}{2}}\rho(r).
\end{equation}
The numerical solution of the above equation is of considerable relevance
to the selected EOS. Numerical solutions of these equations allow
to construct the mass-radius relations of the neutron star (Fig. \ref{rm})
and the gravitational final energy (Fig. \ref{grav}).

\section{\textcolor{blue}{Conclusions}}

The physics of compact objects like neutron stars offers an intriguing interplay
between nuclear processes and astrophysical observables. Neutron stars exhibit
conditions far from those encountered on earth. The determination of an equation
of state (EoS) for dense matter is essential for calculations of neutron star
properties. It is the equation of state that determines the characteristics
and properties of neutron stars such as possible range of masses, the mass-radius
relation and the cooling rate. The astrophysical properties of the compact stars are strongly related
to the properties of quark and nuclear matter. Its mass, radius and
binding energy depend on inner physics of an elementary particle and
nucleus. The linear sigma model predicts properties of the compact
starts which looks quite reasonable. Only the quark star seems to
look as an unphysical object (the positive binding energy, Fig. \ref{grav}).
The astronomical observation of 
a mass - radius relation (for example the SAX J1808.4-3658 \cite{li} compact star) 
may be used to determine the nuclear equation of state.
The best fit to the newly measured radius and mass for EXO 0748-675 
neutron star \cite{exo}
rather prefers (Fig. \ref{rm}) more traditional FST equation of state \cite{serot}.

\end{document}